\documentclass{article}
\usepackage{hyperref}
\usepackage{xspace,soul}
\usepackage{graphicx}
\usepackage{caption}
\usepackage{subcaption}
\usepackage{verbatim}

\newcommand{\R}{\textsf{R}\xspace}
\newcommand{\md}{\textsf{markdown}\xspace}
\newcommand{\RM}{\textsf{R Markdown}\xspace}
\newcommand{\RStudio}{\textsf{RStudio}\xspace}
\newcommand{\knitr}{\textsf{knitr}\xspace}
\newcommand{\cmd}[1]{\texttt{#1}}

\newcommand{\citep}[1]{\cite{#1}}

\begin{document}

\title{R Markdown}
\date{}
\author{Dana Udwin~\thanks{MassMutual Data Labs, Amherst, MA 01027, \texttt{dana.udwin@gmail.com}} \and Ben Baumer~\thanks{Smith College, Northampton, MA 01063, \texttt{bbaumer@smith.edu}}}


\maketitle

\begin{abstract}
Reproducibility is increasingly important to statistical research~\citep{reproducibility}, but many details are often omitted from the published version of complex statistical analyses. A reader's comprehension is limited to what the author concludes, without exposure to the computational process. Often, the industrious reader cannot expand upon or validate the author's results.  Even the author may struggle to reproduce their own results upon revisiting them. \RM is an authoring syntax that combines the ease of Markdown with the statistical programming language R.  An \RM document or presentation interweaves computation, output and written analysis to the effect of transparency, clarity and an inherent invitation to reproduce (especially as sharing data is now as easy as the click of a button).  It is an open-source tool that can be used either on its own or through the \RStudio integrated development environment (IDE)~\citep{r-markdown}.  In addition to facilitating reproducible research, \RM is a boon to collaboratively-minded data analysts, whose workflow can be streamlined by sharing only one master document that contains both code and content.  Statistics educators may also find that \RM is helpful as a homework template, for both ease-of-use and in discouraging students from copy-and-pasting results from classmates. Training students in R Markdown will introduce to the workforce a new class of data analysts with an ingrained, foundational inclination toward reproducible research.
\end{abstract}

The scientific method emphasizes reproducibility as a key component to corroborating and extending results.  While noble in theory, there are roadblocks to realizing reproducibility in statistics and data analysis, chief among them an outdated reliance on copy-and-pasting between computational environments and text editors.  Dividing labor in this way: 1) introduces trivial errors; 2) allows selective reporting; 3) shields the data analysis from public appraisal; 4) impedes intuitive workflow; 5) complicates iterative analysis on new data; 6) makes collaboration awkward; and 7) is time-consuming.  Copying tables and output from one window to another creates opportunity for errors, such as output incongruously placed beside the wrong written analysis.  It also necessitates authorial discretion in deciding which parts of what output are moved into the final report, which can lead to misrepresentation of results.  In the classroom, the copy-and-paste workflow offers a chance to fudge numbers or cop classmates' figures, and may saddle the instructor with a messy, patchwork report to grade.  Peer review, critique, or even further work is not straightforward when the publication does not include code or computation, or includes it in separate, difficult-to-navigate files.

\RM~\citep{markdown} is an open-source authoring format that interweaves written analysis and statistical computation to produce documents, presentations, and other types of reports.  It can be used directly or through the \RStudio IDE, and relies on plain markdown syntax---built on the ethos that a source file should be readable before rendering \citep{nj}---and the statistical programming language \R. A wide array of user-contributed packages, freely available documentation, and a plethora of user-moderated blogs, tutorials, and message boards make \R responsive to users' needs. \RM is equivalently responsive to users' needs.  The code in an \RM document is reevaluated every time the document is rendered, enabling the report to reflect changes in data.  The final output contains code and written analysis where the author(s) wrote it into the source file, as well as output following those generating commands.  Such transparency enhances readers' comprehension and invites review.  Even on the authorial side, integrating computation and textual interpretation creates a natural workflow and smoothes difficulties in collaboration.  If there are multiple authors, they can use online file-sharing services to modify a single \RM document rather than pass back-and-forth dissociated written and computational components that do not respond to changes in the other.  The learning curve with \RM is relatively shallow due to its simple and well-documented syntax.  In short, blending written analysis and statistical computation through \RM is an elegant means to reproducibility.


\section{Overview}

Creating scientific documents is complicated by the necessity of including multiple kinds of information: text, figures, code, and mathematical symbols. \LaTeX~(and its predecessor \TeX) has become the state-of-the-art for scientific papers---due in part to its beautiful and careful rendering of mathematical elements---while Microsoft Word is more commonly used in disciplines that require less mathematical notation. However, neither \LaTeX~nor Word provides the ability to actually compute with data within a document. \RM provides this functionality in a straightforward and easy-to-use manner. 

Like \LaTeX~or HTML, but unlike Word, \RM employs a \emph{source} file and \emph{output} file paradigm. That is, commands and sentences are typed by the user directly into a source file---which is just a plain text file written in a certain format---and this source file is then \emph{rendered} into an output file. The source file typically may only be read by the author, who will then distribute the output file for public consumption.

Yet \RM offers four main advantages over \LaTeX, Word, or HTML for statistical analysis:
\begin{itemize}
	\item Simplicity: Markdown syntax is far simpler than \LaTeX~or HTML. While the text-formatting capabilities are not as feature-rich, they are sufficient for most purposes.
	\item Readability: Markdown syntax is designed so that even the source file is human-readable~\citep{readableMarkdown}. Conversely, \LaTeX~and HTML source documents can be very difficult to parse visually. 
	\item Transparency: All formatting is encoded clearly in a Markdown source file. Conversely, formatting in Word can involve navigating a complex and occasionally inscrutable system of drop-down menus and option windows. 
	\item Embedded Computation: An \RM document contains \R code in its source file, and then the processed results of those commands (along with the commands themselves) in the rendered output file. 
\end{itemize}

Moreover, \RM can produce an output document in either PDF, Word, or HTML format from a single source file. 

Thus, integrating source code, statistical output, and text in \RM is a model of reproducibility.  Such transparency facilitates comprehension, defensibility, and further research or testing. \RM helps to bring the vision for reproducibility in statistical analysis articulated in \cite{gentleman2004statistical} to reality. This vision---in which the barriers to verify another's statistical computations from start to finish are low---is the intellectual descendant of \cite{buckheit1995wavelab} and \citep{claerbout1994hypertext}, and began with Knuth~\cite{knuth1984literate}, the creator of \TeX. Moreover, \RM is dynamic.  Each time the document is rendered, the commands therein are run anew: if data are altered or different data are called in advance of rendering, then the output is dependent and calculated accordingly \citep{regenerate}.

Before returning to more technical details, we give a brief example of how \RM can be used by statistics students for homework assignments (or analysts for reports, et cetera).  

\subsection{An Example Homework Assignment}

\begin{figure}[h]
\centering
\centerline{\includegraphics[width=6in]{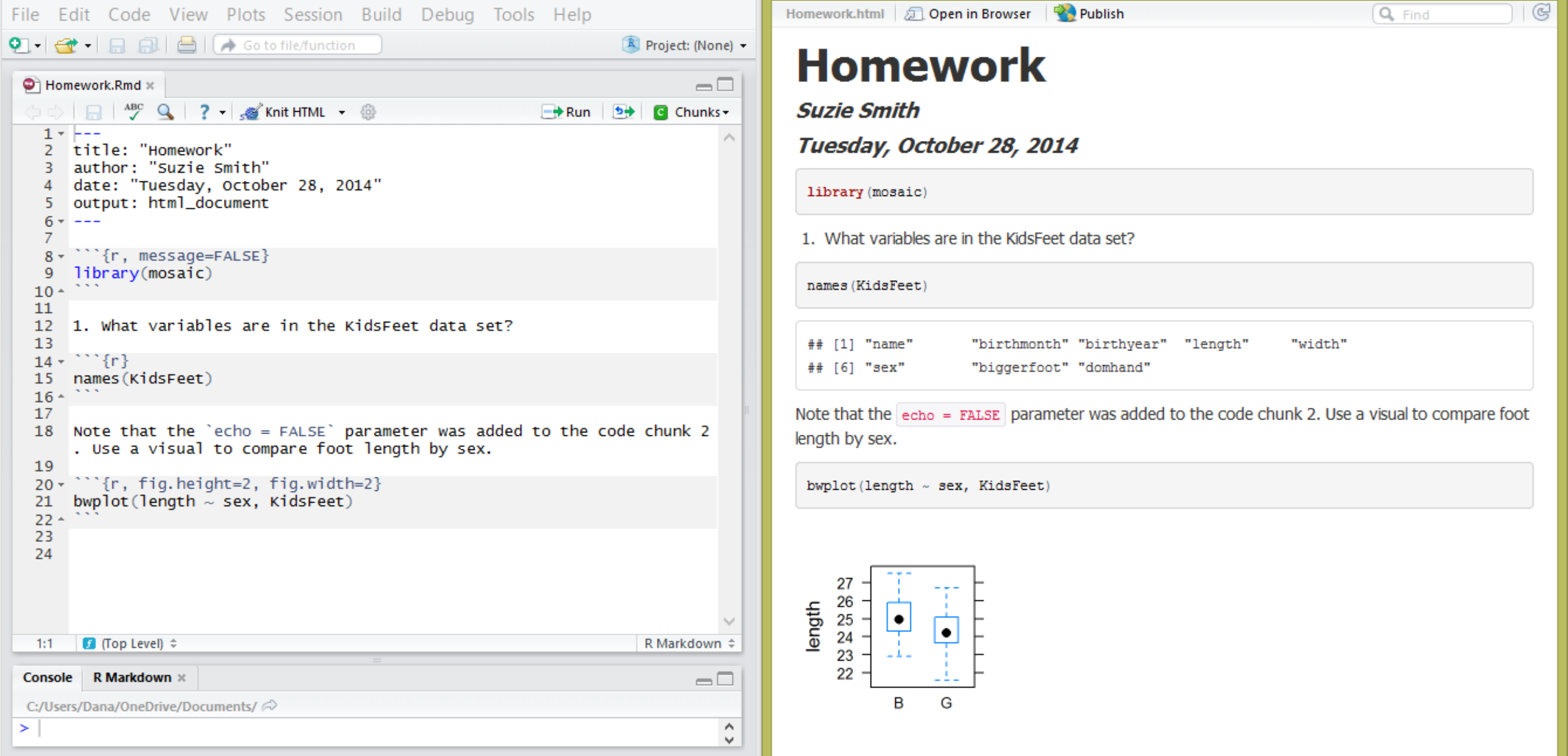}}
\caption{Example of a homework assignment, input (left) and output (right)}
	\label{fig:hw}
\end{figure}

In Figure \ref{fig:hw}, we present an example of how a homework assignment for an introductory statistics student can be written in \RM.  On the left is the \RM source file into which the student would type.  Lines 2-5, sandwiched above and below by three hyphens, contain header information (the syntax is YAML~\citep{yaml}).  All but the \cmd{output: html\_document} is printed in the rendered output shown on the right; this output designation means that the rendered output is an HTML document.  Alternative output formats are PDF and Word.  

The \RM source file contains both written text and ``chunks" of \R code, demarcated by sets of three backticks.  The \RStudio editor shown at left automatically highlights these chunks in light grey.  

Options specific to each chunk may be included in curly backets. For example, \cmd{message=FALSE} omits package-loading messages from the rendered output.  One can see in the rendered output that \RM prints statistical output immediately below the chunk of \R code with which it is associated.

The ease-of-use and transparency inherent to \RM makes it a suitable tool in the introductory statistics classroom.  The alignment of code and output in the final rendered document drives home the connection between the \R commands and the output they create, while eliminating the risk of making a mistake when copy-and-pasting output into a text editor.  Displaying code in the submitted assignment forces students to not simply present a table, model or other output, but also to be able to trace the computational origin of their output.  This may discourage students from cheating.  Students leave the semester with a thorough understanding of the reproducible document and its benefits.  On the educators' side, grading and troubleshooting a single \RM document is less cumbersome than juggling separate written and computational components, such as a Word document and an \R script.

\section{Syntax}

\RM is one of many technologies aiming to provide simple but powerful authoring environments---as opposed to complex, monolithic word-processing applications. Specifically, \md refers to an increasingly popular plain-text authoring syntax designed for simple text documents. For example, when creating installation instructions for a software application, a developer may wish to include simple, functional formatting elements like lists and links, but will likely be reluctant to devote time to curating an elaborate visual appearance for said document (e.g. multiple fonts). \RM is an implementation of \md and includes additional functionality to process output from \R. 

\subsection{Markdown}

In this section we will illustrate some of the simple text formatting features of \md. Note that these have nothing to do with \R. 

\md offers enhanced ease-of-use over other authoring languages like \LaTeX~or HTML.  For example, in order to display the word \textit{data} in italics, we use 

\begin{verbatim}\textit{data}\end{verbatim}

 in \LaTeX,

\begin{verbatim}<i>data<\i>\end{verbatim}

 in HTML, and an uncluttered 
 
\begin{verbatim}*data*\end{verbatim}

 in \md. Additional \md syntax for customizing text is intuitive and straightforward.  

A large heading can be made by underlining a line of text with at least three equal signs. The same construction with hyphens will create a smaller heading. 

\begin{figure}[h]
\centering
\centerline{\includegraphics[width=6in]{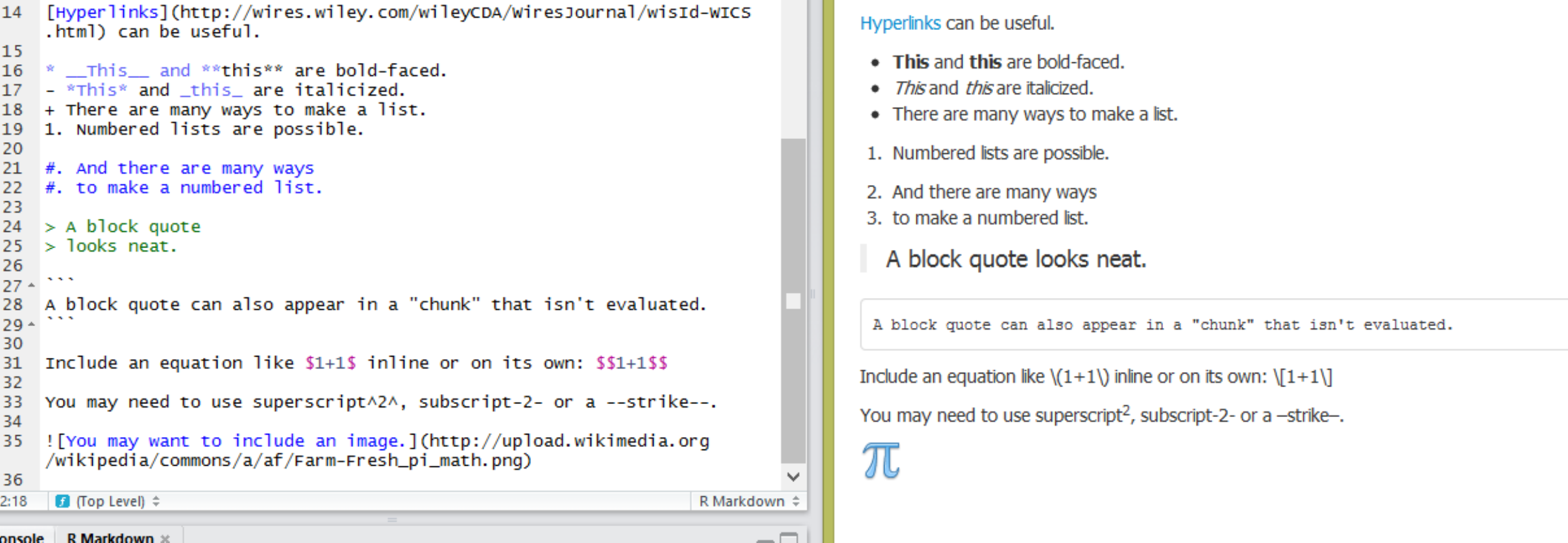}}
\caption{Example of \RM syntax and capabilities, input (left) and output (right)}
	\label{fig:md}
\end{figure}

Formatting commands are equally simple. For example, asterisks, hyphens and plus signs all produce a bulleted list, while either ordinal numbers or pound signs (``\#.") create a numbered list.  A ``greater than" symbol creates a block quote, while typing three backticks above and below text sets the text apart in a fixed-width box (see Figure \ref{fig:md}).

Creating hyperlinks is as easy as including a URL in parentheses---the text that links to the URL immediately precedes the URL in square brackets.  Bold-face and italic text can be created by enclosing that text with two or one asterisks or underscores, respectively.  Carets provide superscripts and tildes enable subscripts or strike-through text.  Images stored locally or remotely (via a URL) can be embedded in the final output as well (see Figure \ref{fig:md}). \RStudio automatically color-codes text in the source file to distinguish between differently formatted text and computation (as shown in Figure \ref{fig:hw}).

\begin{figure}[h]
\centering
\centerline{\includegraphics[width=6in]{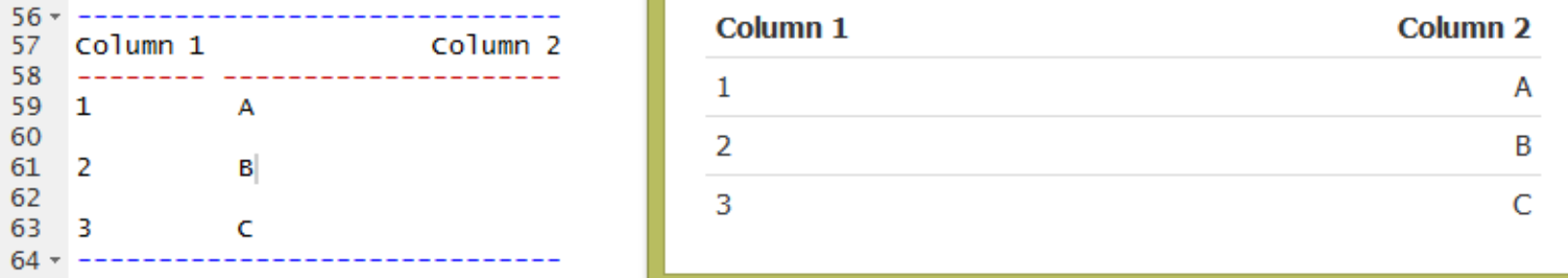}}
\caption{A table created with inline notation, input (left) and output (right)}
	\label{fig:table}
\end{figure}

Tabular information can be typed within appropriately placed strings of hyphens that print as a table in the final rendered output (see Figure \ref{fig:table}).  Tables can be made more complex if necessary.

Although \md does not provide support for all \LaTeX~commands, it can render \LaTeX~equations wrapped in dollar signs using MathJax. More generally, for advanced users who already know HTML, \md will pass chunks of HTML code through to its output. 


\subsection{R + Markdown}

As noted above, \RM provides a particularly valuable extension to \md for statistical analysis, because it enables \R code to be embedded in the source file and rendered as output. There are two ways to include \R code in an \RM document:
\begin{itemize}
	\item Chunks: a block of \R code offset from the main text
	\item Inline: a single line of \R code appearing within the main text
\end{itemize}

\begin{figure}
\centering
\centerline{\includegraphics[width=6in]{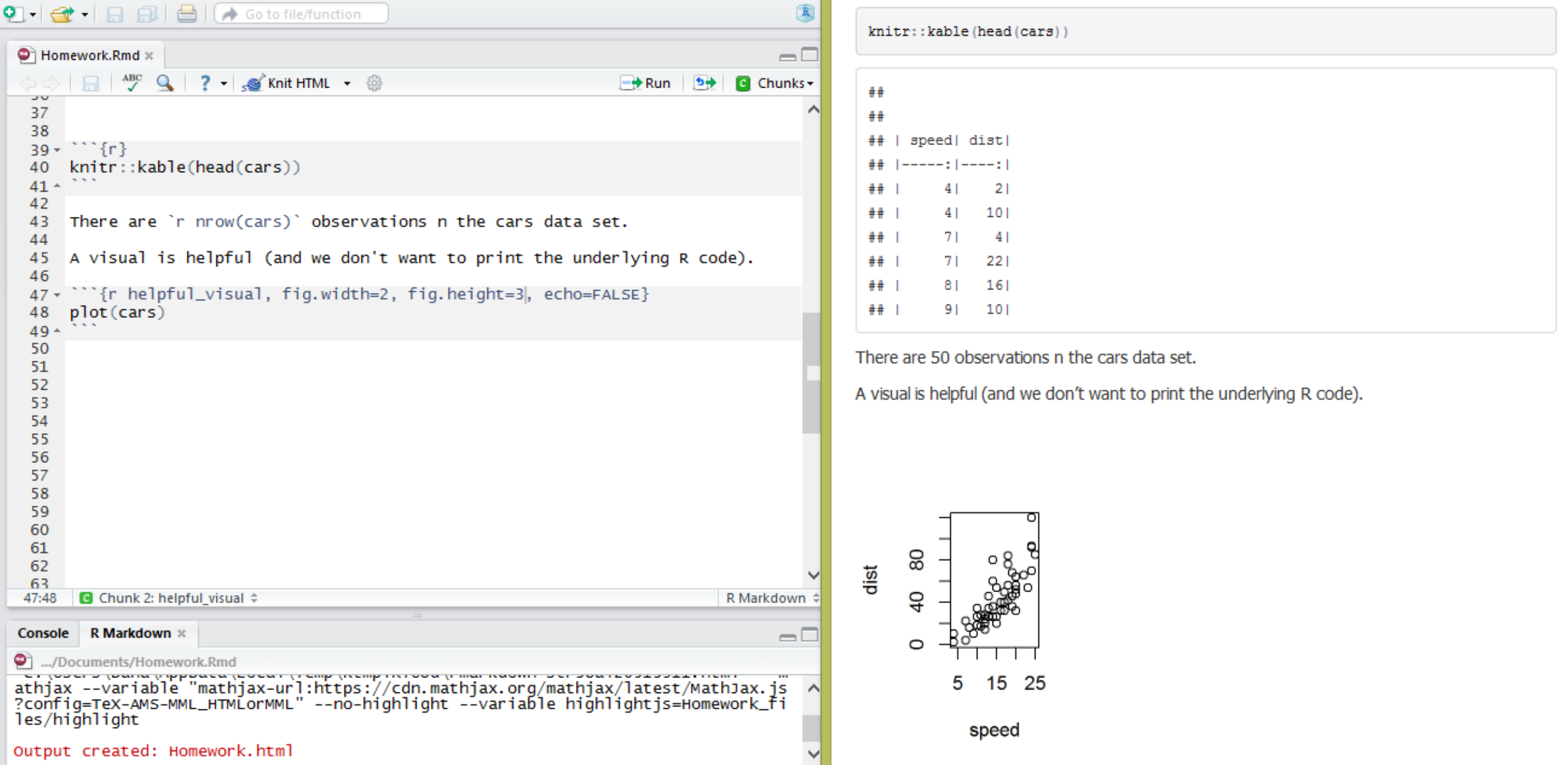}}
\caption{Different uses of an \RM ``chunk'', input (left) and output (right)}
	\label{fig:chunk}
\end{figure}

A chunk---which is executed and printed with the associated computational output when the \RM document is rendered---is created by including three backticks before and after a block of \R code. This is the most common way to incorporate \R commands.  Figure \ref{fig:chunk} shows two separate chunks.  The command in the first chunk is printed in the rendered document.  The second chunk invokes the \cmd{echo=FALSE} option so as not to print the \cmd{plot(cars)} command, although the plot itself still prints.

Alternatively, code can be included in an \RM document inline with text, sandwiched between single backticks.  Inline \R code is evaluated, but not highlighted like ``chunks.''  The first line of text in Figure \ref{fig:chunk} includes \cmd{`r nrow(cars)`}, which evaluates to ``50'' in the final document.

Each \RM document is rendered in a separate, new workspace. Thus, all \R packages as well as data and other objects required for a particular command to be run in the \RM document must be loaded previously in a chunk.  Including \cmd{message=FALSE} in the chunk header suppresses messages generated during evaluations, a useful option when composing an \RM report that requires packages but does not need verbose output (see Figure \ref{fig:hw}). Analogous options exist for \cmd{warnings} and \cmd{errors}. Informing students of this functionality can improve the readability of the rendered output, since many packages produce long, uninformative messages when they are loaded.

There are various other chunk options.  \cmd{include=FALSE} will suppress printing both code and output in the final rendered document, but will still evaluate the chunk when the document is rendered.  \cmd{results='hide'} includes the code but hides the output.  The \cmd{echo=FALSE} option suppresses code but includes output. These options can be useful when, for example, you might want to generate a plot on an exam, but not reveal to students the commands necessary to draw that plot. Or, one can write an exam with embedded solutions, but suppress those solutions with \cmd{echo=FALSE} to generate an exam copy. It is wise to name your chunks (as modeled by the final chunk in Figure 4, called \cmd{helpful\_visual}), so that the error report generated in the case of failed rendering can pinpoint the problematic chunk by name.  Chunk options for plots like \cmd{fig.width} and \cmd{fig.height} set the size for plots created in the chunk, as shown in Figure \ref{fig:hw} and Figure \ref{fig:chunk}.


Often, an author will desire uniform chunk options throughout their report.  In this case, there are \R commands to specify global chunk options, which one might include in a chunk that \cmd{include=FALSE} hides in the final rendered document. It is also possible to defer output to the end of a document, as a means of automatically creating a technical appendix.

\subsection{Nuts and Bolts}

\RM---which was developed by the \RStudio team---works in conjunction with the \knitr package~\citep{knitr}---which was created by Yihui Xie as part of his graduate work at Iowa State University~\citep{xie2013knitr}. \knitr is the successor of Sweave~\cite{leisch2002sweave}. Upon his graduation, Xie was hired by \RStudio, and is now the maintainer of the \RM package. Thus, while \R, \md, \RM, and \knitr were developed separately by independent groups of people, \RStudio now quite deliberately maintains the \RM portion of this universe. In this case, the result is a seamless integration of \RM functionality---using \knitr---in \RStudio. 

As noted above, \md is a general-purpose text authoring format for documents, and \RM is simply \md, with functionality for computing with \R injected into it. \knitr is the rendering engine that converts an \RM document into HTML. In fact, \knitr is capable of much more, including rendering Sweave documents that combine \LaTeX~and \R code. The extensive leveraging of existing technologies imbues \RM with powerful functionality. The current verson of \RM uses Pandoc~\citep{pandoc}---an all-purpose text file conversion program---to dramatically increase its versatility and long-term viability. For example, it is now possible to make an entire modern-looking website (such as the \RM website itself) using \RM and writing just a few lines of HTML code. 

\section{Workflows}

In this section, we discuss some common situations in which workflows may be significantly streamlined through the adoption of \RM. 

\subsection{Student Homework}

Instructors at several institutions have found \RM to be a useful tool for student homework in both introductory and higher-level statistics courses~\citep{baumer2014r}. Having a student's work---both computational and analytical---in one document provides benefits to both the student and their instructor. 
The transparency created by integrating code and written analysis simplifies troubleshooting and dissuades cheating.  Underlying \RM is an emphasis on reproducibility that students may carry forward in future coursework and careers.

\RM also subverts the temptation to copy-and-paste statistical output from the computation environment into a text editor like Word.  Flitting from window to window increases the likelihood of errors, such as misaligning statistical output with the wrong exposition.  A less benign consequence of the copy-and-paste workflow is an increase in the number o f opportunities for students to selectively report results.  \RM mitigates the risk of selective reporting by requiring all code to be contained within and printing all output by default.  In addition, weaving together statistical computation and written analysis is intuitive in a way that mirrors statisticians' process and betters readers' comprehension.

\subsection{Collaborative Research}

As a well-documented open-source programming language, \R itself is already an emblem of accessibility. Moreover, we have argued that publishing code, output, and written analysis in an integrated document fosters collaboration and substantive critique.  A variety of blogs and forums host active information exchanges for debugging and development.  Designing a package specific to one's needs (there are 5,696 packages at last count~\citep{toomany}) is a characteristic activity for an advanced \R user. Many \R projects employ the version-control system Git through the web-sharing service GitHub to facilitate group development~\citep{git}. Indeed, GitHub uses GitHub-flavored \md for its text files, so \RM users will feel right at home in that system. More generally, \RM belongs to the same ethos of collaborative development. (In fact, the \md package for \R is hosted on GitHub!)

More advanced features are available to \RM users who desire publication-quality appearance. 
One of these nearly 6,000 \R package, \cmd{xtable}, coerces data to \LaTeX or HTML tables. Citations---another crucial component of fully transparent research---are easily incorporated by referencing a bibliography file in the \RM document's title header (i.e. \cmd{bibliography: bib\_name.bib}).  To cite an entry from the bibliography, type [$@$item\_ID] in the \md text environment, where \cmd{item\_ID} is the citation identifier in the bibliography.

Statistical analysis is increasingly the work of teams, and \RM facilitates collaboration within these teams by keeping the narrative and computation inherent in data analysis together in a single place. Functionality provided by \textsf{packrat}~\citep{packrat}---yet another contribution from the \RStudio team---enables seamless package version management. 

\subsection{RStudio Integration}

Although \RM is developed by \RStudio, one can use \RM and \knitr outside of the \RStudio IDE. However, users of \RStudio will find several tightly integrated features for working with \RM.

\begin{figure}
\centering
\centerline{\includegraphics[width=4in]{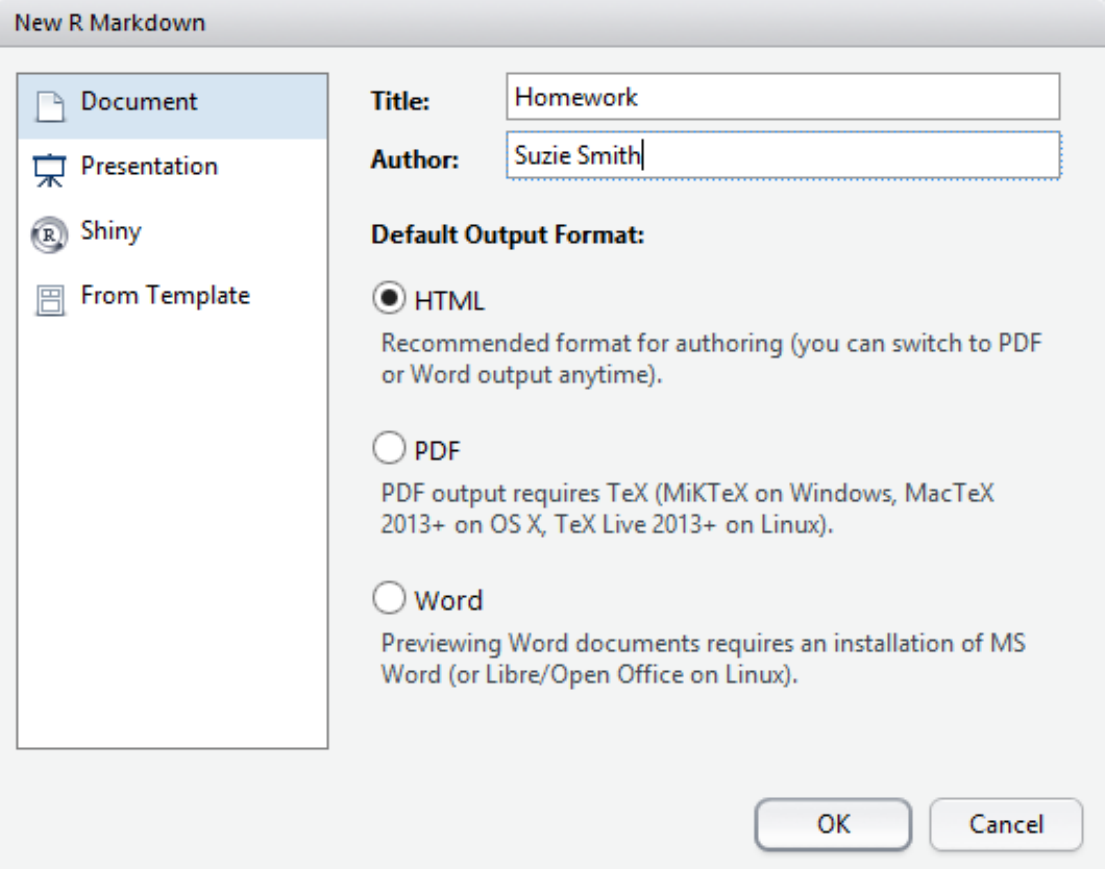}}
\caption{\RM dialog box to create a new document, input (left) and output (right)}
	\label{fig:dialog}
\end{figure}

\RM is among the file types listed in the \RStudio File $\rightarrow$ New File drop-down menu (see Fig.~\ref{fig:dialog}).  After electing to build an \RM document in \RStudio, the user may choose an output type (document, presentation or Shiny) or rely on a template.  Templates are typically found within the \cmd{inst/rmarkdown/templates} directory of an \R package and are an opportunity to create content using a standardized format. Some popular formats (e.g. the \textit{Journal of Statistical Software} template in the \cmd{rticles} package \citep{rticles}) are publicly available, but users are free to create their own.

\RStudio offers additional flexibilty after the author has built their \RM source file and is ready to render the document.  The ``knit'' button has a drop-down option that offers different output options depending on if the user is building a document, a presentation or a Shiny web application. 

\section{Output Formats}

If using \RM in the \RStudio IDE, then the document is rendered by clicking a button labeled ``Knit HTML"; equivalently, calling the:
\begin{verbatim}
	rmarkdown::render('doc.rmd')
\end{verbatim}
function outside \RStudio does the job.  

The current version of \RM is based on Pandoc as well as \knitr, and can therefore produce any of HTML, PDF, and Word document types.  \cmd{Beamer}, \cmd{ioslides}, and \cmd{reveal.js} presentations are also possible.  Interactivity has been introduced through integration with Shiny, a web application framework with \R.

In what follows, we outline the three major types of output documents. 

\subsection{Document}

The most common use for \RM is to generate static documents.

\begin{figure}
\centering
\centerline{\includegraphics[width=6in]{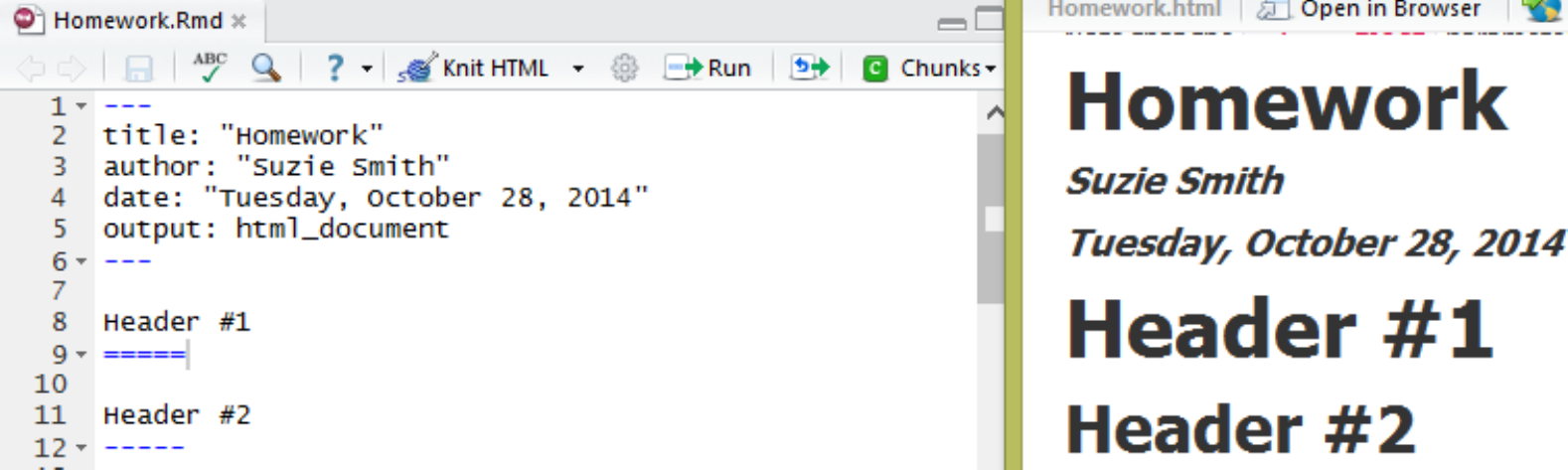}}
\caption{Example of a YAML document header, input (left) and output (right)}
	\label{fig:header}
\end{figure}

After filling in the empty fields and selecting an output format from HTML, PDF, or Word in the dialog box (Fig.~\ref{fig:dialog}), a new document with a pre-filled YAML header is created (see Figure~\ref{fig:header}).  Several additional parameters can be added to the header, such as table of contents (``toc:''), themes, or output formats which can be rendered simultaneously.  For maintaining a consistent look-and-feel across multiple documents (e.g. a website), the YAML header may be saved as a separate file. Any \RM documents in the same directory will automatically use this header information. 

In addition, the latest version of \RM allows for footnotes and citations with comparable ease.

\subsection{Presentation}

\RM can generate presentations in either HTML (ioslides, reveal.js) or PDF (Beamer) format. The former takes advantage of newer features in HTML5, and can be viewed in any modern web browser. Conversely, the latter requires a local installation of \TeX. Similar to documents, a user can seamlessly render the same \RM presentation in either output format, without changing the content of the source file. 

The format for \RM presentations is similar to that of documents, with individual slides demarcated by double pound signs (\#\#). As with \RM documents, the YAML header controls various options, including the overall display size (e.g. widescreen), text size, bullet formatting, and transition speed.  The header will also take a logo option so that an image of the author's choice is projected onto the title page and slide footers.  Options like \cmd{fig\_height} and \cmd{fig\_width} control the default figure size through the presentations.

\begin{figure}
\centering
\centerline{\includegraphics[width=6in]{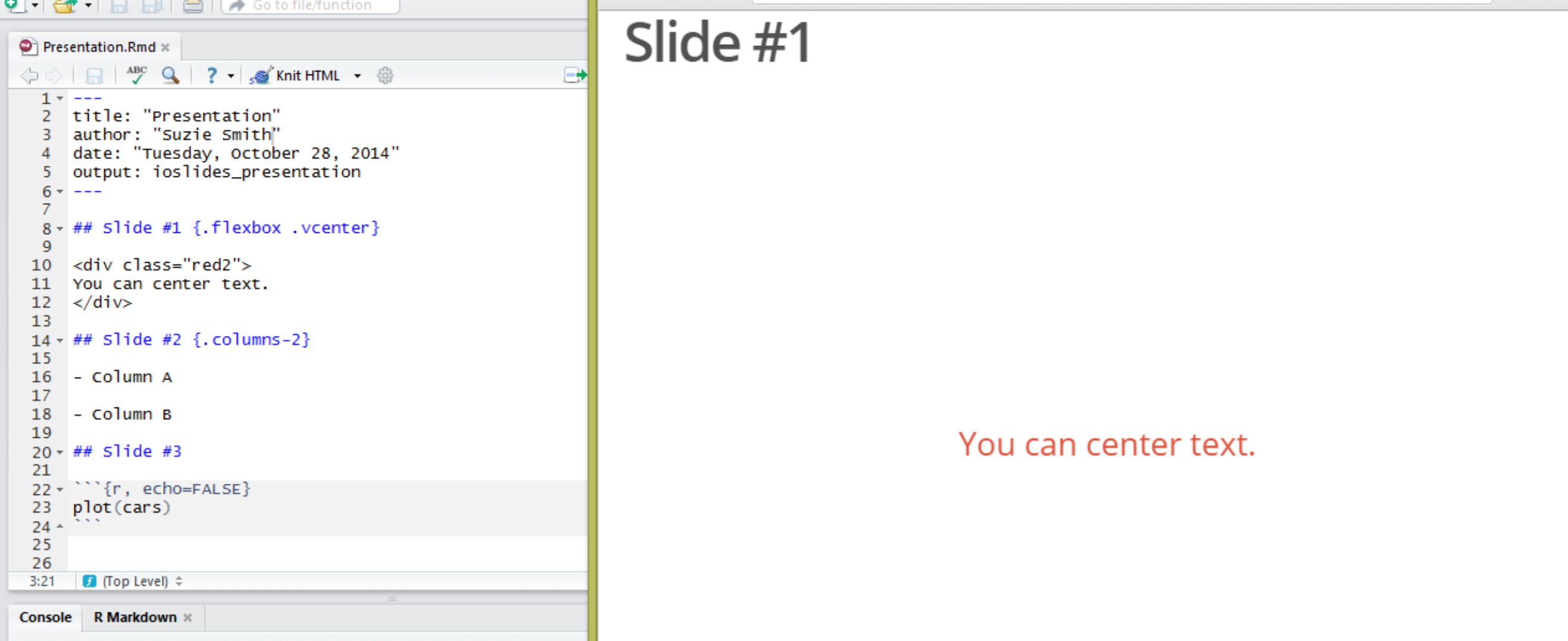}}
\caption{\RM ioslide presentation, input (left) and output for Slide 1 (right)}
	\label{fig:sourceio}
\end{figure}

\begin{figure}
\centering
\begin{subfigure}[b]{0.24\textwidth}
\includegraphics[width=\textwidth]{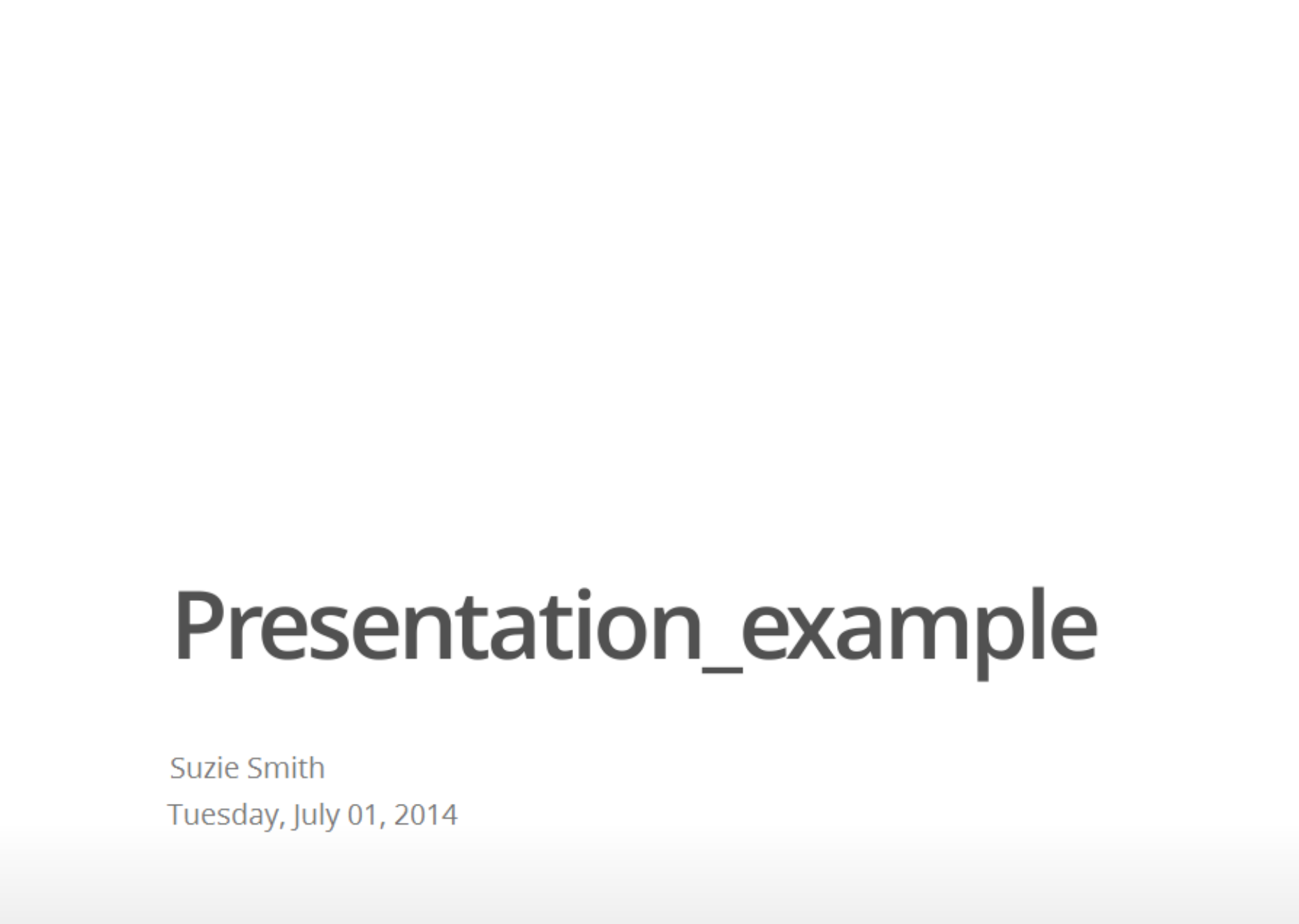}
\end{subfigure}
\begin{subfigure}[b]{0.24\textwidth}
\includegraphics[width=\textwidth]{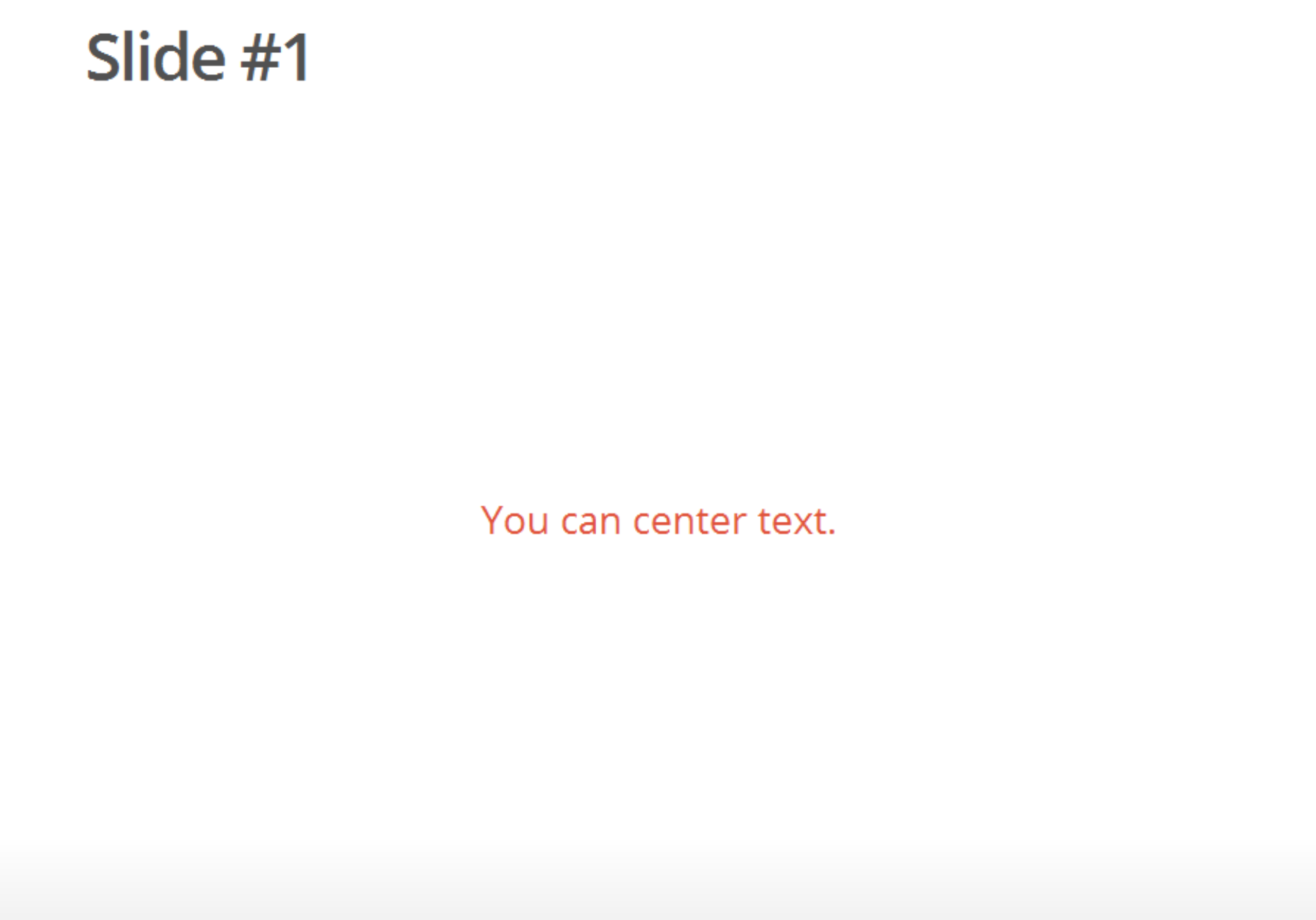}
\end{subfigure}
\begin{subfigure}[b]{0.24\textwidth}
\includegraphics[width=\textwidth]{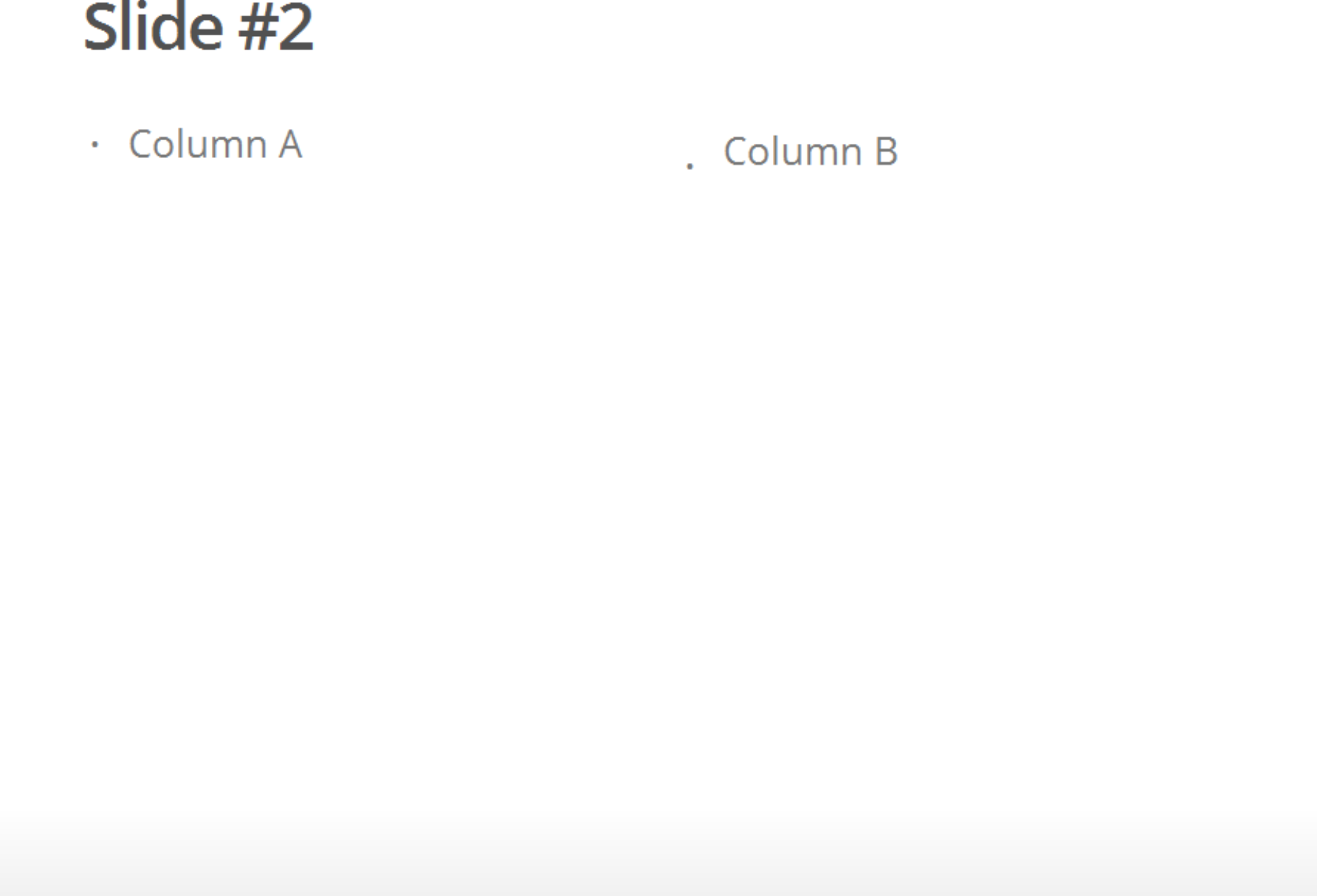}
\end{subfigure}
\begin{subfigure}[b]{0.24\textwidth}
\includegraphics[width=\textwidth]{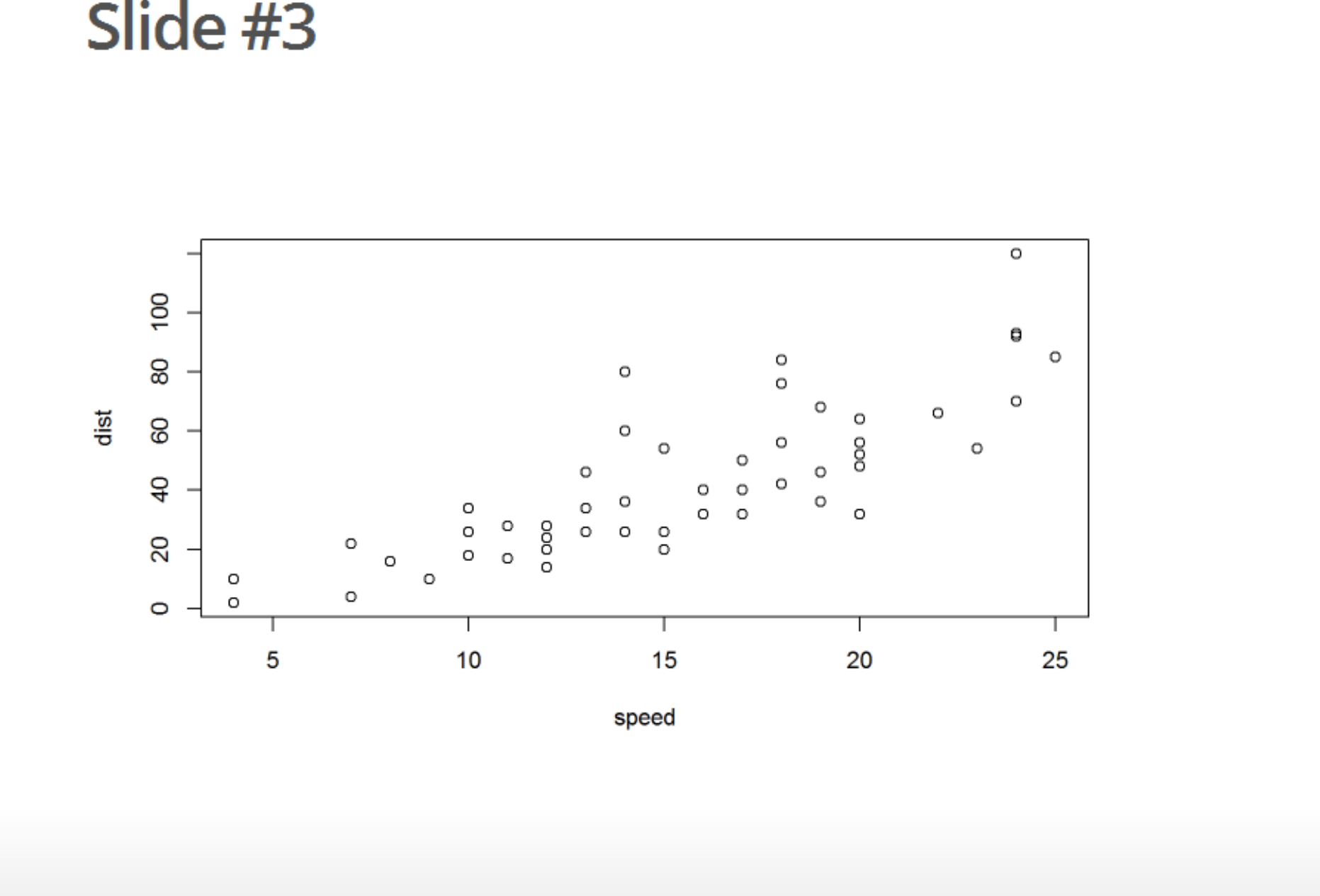}
\end{subfigure}
\caption{Rendered ioslides}
	\label{fig:evalio}
\end{figure}

Outside the header, two pound signs signifies the start of a new slide.  CSS attributes---which can also be defined in an external file---in curly brackets follow the slide title.  In Figure \ref{fig:sourceio}, the ``.flexbox" and ``.vcenter" attributes enable center-aligned text.  Calling \cmd{class="red2"} creates colored text. 
Bulleted lists mimic the syntax in an ordinary \RM environment, as do chunk options and \R commands.

\subsection{Shiny}

Shiny is a web application framework for \R that can be leveraged in \RM to enable interactivity. 
As opposed to the document and presentation outputs, which are static, a Shiny application is dynamic, and contains elements
that the reader can manipulate.

\begin{figure}
\centering
\centerline{\includegraphics[width=6in]{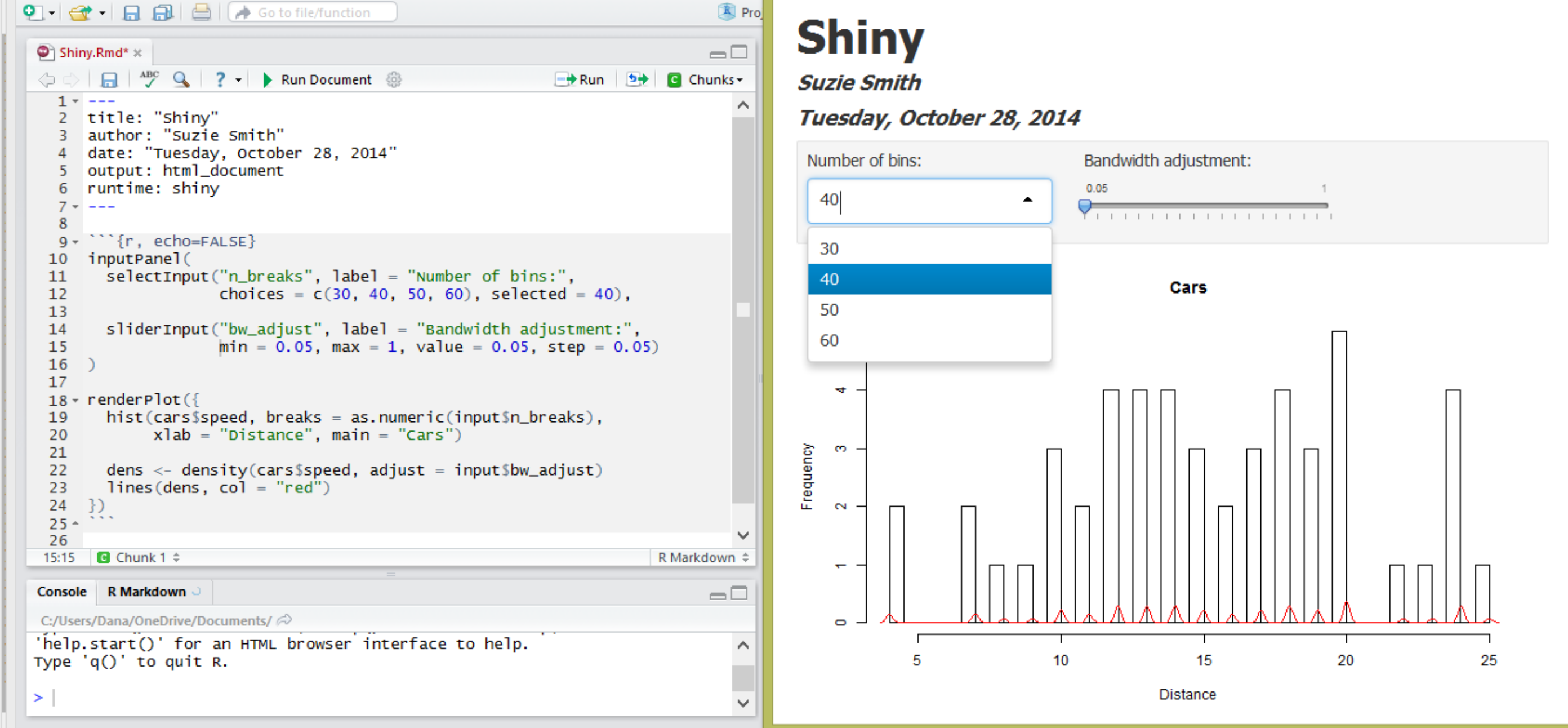}}
\caption{Shiny code that generates an interactive plot, input (left) and output (right)}
	\label{fig:renderplot}
\end{figure}

Shiny code can be used to generate dynamic plots, as demonstrated in Figure \ref{fig:renderplot} with the \cmd{renderPlot()} function.  The histogram and density plots called in \cmd{renderPlot()} rely on user input parameters defined in the \cmd{inputPanel()} function.

\begin{figure}
\centering
\centerline{\includegraphics[width=6in]{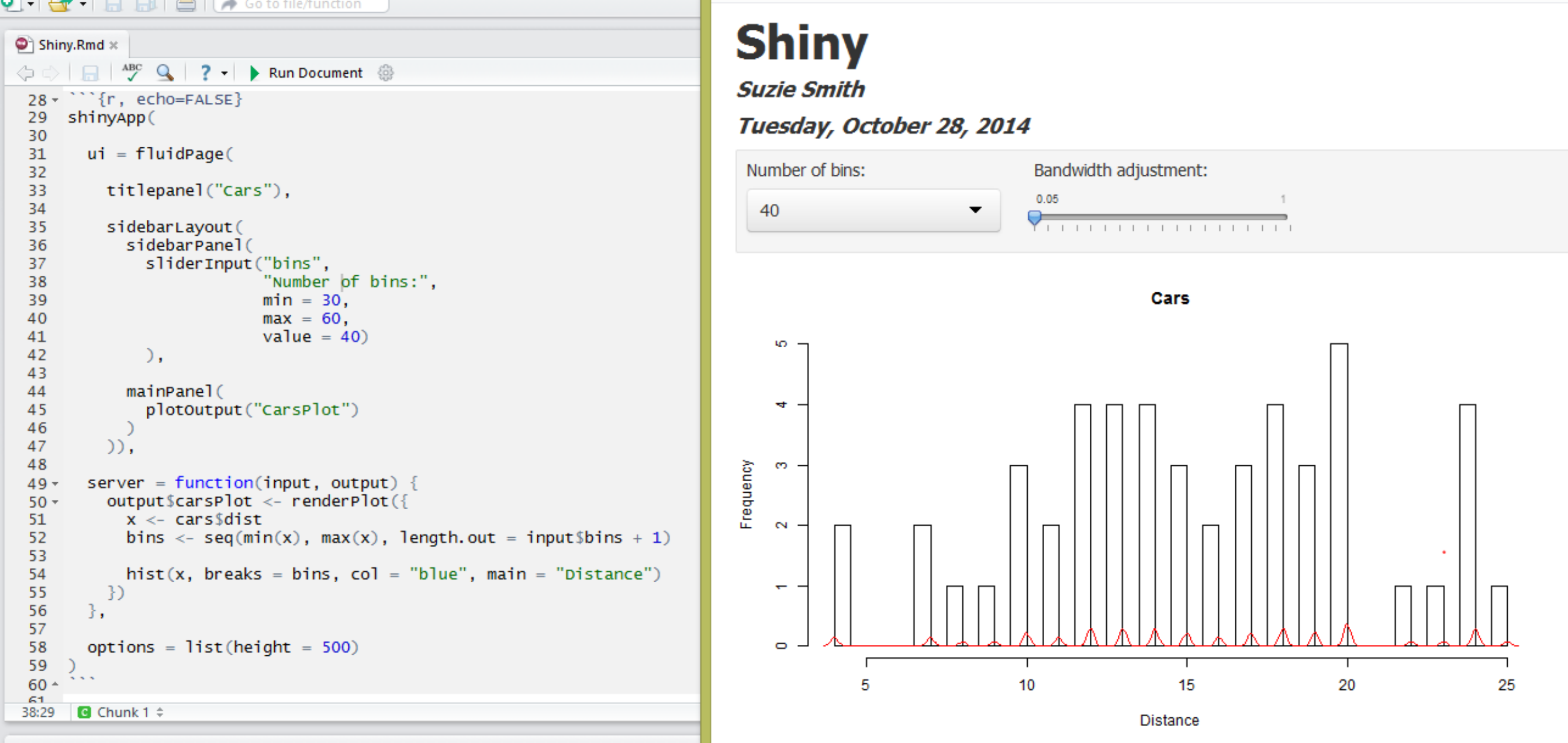}}
\caption{Full Shiny application defined within an \RM chunk, input (left) and output (right)}
	\label{fig:shinyapp}
\end{figure}

Alternatively, an entire Shiny application can be embedded in the document, by either defining it within a chunk using the \cmd{shinyApp()} function or calling a Shiny application defined elsewhere using the \cmd{shinyAppDir()} function.  Figure \ref{fig:shinyapp} shows \cmd{shinyApp()} at work. 


%

\section{Conclusion}

\RM is a statistical authoring tool that meets high standards of usability, reproducibility, and functionality. For conducting research and assembling a report, \RM offers an intuitive workflow that allows for interspersed computation and written analysis in a single source file.  Combining statistical programming and text in one platform eases the strain of collaboration when there is more than one researcher at work.  Because the \R code is reevaluated every time an \RM document is rendered modifications to the data are automatically reflected in the output of subsequent renderings. Statisticians with changing data need no longer spend time recoding and copy-and-pasting output.  They also have increased flexibility for sharing their work: presentation or document, static, or with dynamic user-controlled elements, with all the text-formatting customization available in \md.  In terms of possible data manipulation, visualization and general computational capabilities, their arsenal is as expansive as \R: that is, virtually limitless, as \R is open-source with a growing number of packages.  When the paper or presentation is complete and the final product distributed, readership can survey work that is transparent, with potentially every table, figure and computation matched to the \R code responsible.  The discriminating reader can trust that the work is sound (or diagnose missteps in approach) and may be empowered to reproduce the research.  Publishing computational process alongside results facilitates academic dialogue.  

\RM is also a natural fit in the statistics classroom (introductory or advanced), where using one platform to import, clean, analyze, and interpret data provides a streamlined workflow.  Submitting one document for grading in turn simplifies the instructor's work, who would otherwise review isolated output without specific information as to what went wrong in calculation.  When students compile a semester's work to study for exams, they will review documents that clearly associate code and output.  In short, \RM satisfies the call for reproducibility in scientific research while also improving workflow.


\bibliographystyle{plain}
\bibliography{references}


\end{document}